\documentclass[aps,twocolumn,groupedaddress,showpacs]{revtex4}
\usepackage[dvips]{graphicx}
\usepackage[]{caption}
\usepackage{amsmath}
\usepackage{amssymb}
\def\Dsl{\hbox{/\kern-.6700em\it D}} 
\def\dsl{\hbox{/\kern-.5300em$\partial$}}

\def\eqa{\begin{eqnarray}}
\def\eeqa{\end{eqnarray}}
\def\eq{\begin{equation}}
\def\eeq{\end{equation}}
\def\be{\begin{equation}}
\def\ee{\end{equation}}
\def\bea{\begin{eqnarray}}
\def\eea{\end{eqnarray}}
\def\nn{\nonumber}

\def\v{ \vec}

\begin{document}
\bibliographystyle{prsty}
\title{Back-Reaction: A Cosmological Panacea}
\author{ P. Martineau and R. Brandenberger}
\affiliation{Dept. of Physics, McGill University, 3600 University Street,
Montr\'eal QC, H3A 2T8, Canada }
\date{\today}
\pacs{98.80.Cq}
\begin{abstract}
We present a solution to the dark energy problem in terms of the 
Effective Energy Momentum Tensor (EMT) of cosmological perturbations. 
This approach makes use of the 
gravitational back-reaction of long wavelength (super-Hubble) fluctuation 
modes on the background metric. Our results indicate that, following 
preheating, the energy density associated with back-reaction is sub-dominant 
and behaves as a tracker during the radiation era. At the onset of matter 
domination, however, the effects of back-reaction begin to grow relative to 
the matter density and the associated equation of state quickly approaches 
that of a cosmological constant. Using standard values for the preheating 
temperature and the amplitude of the inflaton following preheating, we 
show that this mechanism leads to a very natural explanation of dark energy. 
We comment on other recent attempts to explain the dark energy using back-reaction and their relation to our work.
\end{abstract}
\maketitle

\section{Introduction}

The nature and origin of dark energy stand out as two of the great 
unsolved mysteries of cosmology. Two of the more popular explanations are 
either a cosmological constant $\Lambda$, or a new, slowly rolling scalar
field (a quintessence field). If the solution of the dark energy problem 
proved to be a cosmological constant, one would have to explain why it is not
120 orders of magnitude larger (as would be expected in a non-supersymmetric
field theory), nor exactly zero (as it would be if some hidden symmetry
were responsible for the solution of the cosmological constant problem), and 
why it has become dominant only recently in the history of the universe. These
are the ``old'' and ``new'' cosmological constant problems in the parlance 
of \cite{Weinberg:2000yb}. To date, this has not been accomplished 
satisfactorily, despite intensive efforts. If, instead of $\Lambda$, the 
solution rested on quintessence, one would need to justify the existence of 
the new scalar fields with the finely tuned properties required of a
quintessence field (e.g. a tiny mass of about $10^{-33}$eV if the field
is a standard scalar field). Clearly, both of the above approaches
to explaining dark energy lead directly to serious, new cosmological problems.
In this paper, we will explore an approach to explaining dark energy which
does not require us to postulate any new matter fields.

There exist tight constraints on $\Lambda$ from various sources -  
Big Bang Nucleosynthesis (BBN) \cite{Freese:1986dd}, cosmic microwave
background (CMB) anisotropies \cite{Bean:2001xy}, cosmological structure 
formation \cite{Doran:2001rw} - which rule out models where the vacuum 
energy density is comparable to the matter/radiation energy density at the
relevant cosmological times in the past. However, it could still be hoped 
that a variable $\Lambda$ model might be compatible with observation since 
the value of $\rho_{\Lambda}$ is constrained only for certain redshifts. 
In fact, the above constraints taken together with the results from recent 
supernovae observations \cite{Riess:1998cb},\cite{Perlmutter:1998np} leads 
one to posit that the vacuum energy density might be evolving in time. 

This leads directly to the proposal of tracking quintessence 
\cite{Ratra:1987rm}. However, some of the drawbacks of quintessence were 
mentioned above. A preferable solution would combine the better features of 
both quintessence and a cosmological constant: a tracking cosmological 
``constant''.

In this letter, we discuss the possibility that the energy-momentum
tensor of long wavelength cosmological perturbations might provide
an explanation of dark energy. The role of such perturbations in
terminating inflation and relaxing the bare cosmological constant
was investigated some time ago in \cite{Mukhanov:1996ak,Abramo:1997hu} (see
also \cite{WT}). However, this mechanism can only set in if the
number of e-foldings of inflation is many orders of magnitude larger
than the number required in order to solve the horizon
and flatness problems of Standard Big Bang cosmology. Here, we are
interested in inflationary models with a more modest number of
e-foldings. We discover that, in this context, the EMT of long wavelength
cosmological perturbations results in a tracking cosmological ``constant'' 
of purely gravitational origin and 
can be used to solve the ``new'' cosmological constant problem. 

We begin by reviewing the formalism of the effective EMT of cosmological
perturbations in Section 2. We recall how, in the context of
slow-roll inflation, it could solve the graceful exit problem of
certain inflationary models. We then extend these results beyond the 
context of slow-roll inflation in Section 3. In Section 4, we investigate 
the behaviour of the EMT during the radiation era and show that the 
associated energy density is sub-dominant and tracks the cosmic fluid. 
We examine the case of the matter era and show how the EMT 
can solve the dark energy problem in section 5. In Section 6 we consider the
effects of back-reaction on the scalar field dynamics. We then 
summarize our results and
comment on other attempts to use the gravitational back-reaction of
long wavelength fluctuations to explain dark energy.

\section{The EMT}

The study of effective energy-momentum tensors for gravitational 
perturbations is not new \cite{Brill,Isaacson}. The interests of 
these early authors revolved around the effects of high-frequency 
gravitational waves. More recently, these methods were applied
\cite{Mukhanov:1996ak,Abramo:1997hu} to the study of the effects of 
long-wavelength scalar metric perturbations and its application to 
inflationary cosmology.

The starting point was the Einstein equations in a background defined by
\bea
ds^{2} \, &=& \, a^{2}(\eta)((1+2\Phi(x,\eta))d{\eta}^{2}\nn \\ 
&-&(1-2\Phi(x,\eta))(\delta_{ij}dx^{i}dx^{j}))
\eea
where $\eta$ is conformal time, $a(\eta)$ is the cosmological
scale factor, and $\Phi(x, \eta)$ represents the scalar perturbations (in
a model without anisotropic stress). We are using longitudinal gauge
(see e.g. \cite{MFB} for a review of the theory of cosmological fluctuations,
and \cite{RHBrev03} for a pedagogical overview). Matter is, for simplicity,
treated as a scalar field $\varphi$.

The modus operandi of \cite{Mukhanov:1996ak} consisted of expanding both the 
Einstein and energy-momentum tensor in metric ($\Phi$) and matter 
($\delta\varphi$) perturbations up to second order. The linear equations 
were assumed to be satisfied, and the remnants were spatially averaged, 
providing the equation for a new background metric which takes into
account the back-reaction effect of linear fluctuations computed up to
quadratic order
\be
G_{\mu\nu} \, = \, 8\pi G\,[T_{\mu\nu}+\tau_{\mu\nu}],
\ee
where $\tau_{\mu\nu}$ (consisting of terms quadratic in metric and matter
fluctuations) is called the effective EMT.

The effective energy momentum tensor, $\tau_{\mu\nu}$, was found to be
\begin{eqnarray}  \label{tzero}
\tau_{0 0} &=& \frac{1}{8 \pi G} \left[ + 12 H \langle \phi \dot{\phi} \rangle
- 3 \langle (\dot{\phi})^2 \rangle + 9 a^{-2} \langle (\nabla \phi)^2
\rangle \right]  \nonumber \\
&+& \,\, \langle ({\delta\dot{\varphi}})^2 \rangle + a^{-2} \langle
(\nabla\delta\varphi)^2 \rangle  \nonumber \\
&+& \,\,\frac{1}{2} V''(\varphi_0) \langle \delta\varphi^2 \rangle + 2
V'(\varphi_0) \langle \phi \delta\varphi \rangle \quad ,
\end{eqnarray}
and
\begin{eqnarray}  \label{tij}
\tau_{i j} &=& a^2 \delta_{ij} \left\{ \frac{1}{8 \pi G} \left[ (24 H^2 + 16 
\dot{H}) \langle \phi^2 \rangle + 24 H \langle \dot{\phi}\phi \rangle
\right. \right.  \nonumber \\
&+& \left. \langle (\dot{\phi})^2 \rangle + 4 \langle \phi\ddot{\phi}\rangle
- \frac{4}{3} a^{-2}\langle (\nabla\phi)^2 \rangle \right] + 4 \dot{{%
\varphi_0}}^2 \langle \phi^2 \rangle  \nonumber \\
&+& \,\, \langle ({\delta\dot{\varphi}})^2 \rangle -  a^{-2} \langle
(\nabla\delta\varphi)^2 \rangle - 
4 \dot{\varphi_0} \langle \delta \dot{\varphi}\phi \rangle  \nonumber \\
&-& \left. \,\, \, \frac{1}{2}V''(\varphi_0) \langle \delta\varphi^2
\rangle + 2 V'( \varphi_0 ) \langle \phi \delta\varphi \rangle
\right\} \quad ,
\end{eqnarray}
where H is the Hubble expansion rate and the $\langle \rangle$ denote 
spatial averaging.

Specializing to the case of slow-roll inflation (with $\varphi$ as the
inflaton) and focusing on the effects of long wavelength or IR modes
(modes with wavelength larger than the Hubble radius), the EMT simplifies to
\be
\tau _0^0 \cong \left( 2\,{\frac{{V^{\prime \prime }V^2}}
{{V^{\prime }{}^2}}}-4V\right) <\phi ^2> \, \cong \, \frac 13\tau_i^i, 
\ee 
and
\be 
p \, \equiv -\frac 13\tau _i^i\cong -\tau_{0}^{0}\,.  
\ee
so that $\rho_{eff}<0$ with the equation of state $\rho\,=\,-p$.

The factor $\langle \phi^{2} \rangle$ is proportional to the IR phase space 
so that, given a sufficiently long period of inflation (in which the phase 
space of super-Hubble modes grows continuously), $\tau_{0}^{0}$ can become 
important and act to cancel any positive energy density (i.e. as associated 
with the inflaton, or a cosmological constant) and bring inflation to an end 
- a natural graceful exit, applicable to any model in which inflation 
proceeds for a sufficiently long time.

Due to this behaviour during inflation, it was speculated 
\cite{Brandenberger:1999su} that this could also be used as a mechanism 
to relax the cosmological constant, post-reheating - a potential solution to 
the old cosmological constant problem. However, this mechanism works (if
at all - see this discussion in the concluding section) only if inflation
lasts for a very long time (if the potential of $\varphi$ is quadratic,
the condition is that the initial value of $\varphi$ is larger than
$m^{-1/3}$ in Planck units).

\section{Beyond Slow-Roll}

Here, we will ask the question what role back-reaction of IR modes
can play in those models of inflation in which inflation ends naturally 
(through the reheating dynamics of $\varphi$) before the phase space of 
long wavelength modes has time to build up to a dominant value.
In order to answer this question, we require an expression for 
$\tau_{\mu\nu}$ unfettered by the slow-roll approximation. Doing this provides 
us with an expression for the EMT which is valid during preheating and, 
more importantly, throughout the remaining course of cosmological evolution.

In the long wavelength limit, we have \footnote{We've ignored terms proportional to $\dot{\phi}$ on the 
basis that such terms are only important during times when the equation of 
state changes. Such changes could lead to large transient effects during 
reheating but would be negligible during the subsequent history of the 
universe.},
\begin{eqnarray}  \label{one}
\tau_{0 0} &=& \frac{1}{2} V''(\varphi_0) \langle \delta\varphi^2 \rangle + 2
V'(\varphi_0) \langle \phi \delta\varphi \rangle, \quad 
\end{eqnarray}
and 
\begin{eqnarray}  \label{two}
\tau_{i j} &=& a^2 \delta_{ij} \left\{ \frac{1}{8 \pi G} \left[ (24 H^2 + 16 
\dot{H}) \langle \phi^2 \rangle] + 4 \dot{{\varphi_0}}^2 \langle \phi^2 \rangle\} \right.  \right. \nonumber \\
 &-& \left. \,\, \, \frac{1}{2}V''(\varphi_0) \langle \delta\varphi^2
\rangle + 2 V'( \varphi_0 ) \langle \phi \delta\varphi \rangle
\right\}.
\end{eqnarray}

As in the case of slow-roll, we can simplify these expressions by making use 
of the constraint equations which relate metric and matter
fluctuations \cite{MFB}, namely
\be \label{constr}
-(\dot{H} + 3H^2) \phi \, \simeq \, 4 \pi G V^{'} \delta \varphi \, .
\ee
Then, (\ref{one}) and (\ref{two}) read
\be{\tau_{0 0}\,=\, (2\kappa^2\frac{V''}{(V')^{2}}(\dot{H}+3H^{2})^{2}-4\kappa(\dot{H}+3H^{2}))\langle \phi^{2}\rangle,}\ee\label{1}
\begin{eqnarray}
\tau_{i j}\,&=&\, a^{2}\delta_{i j}(12\kappa(\dot{H}+H^{2})+4\dot{\varphi_{0}(t)}^{2}\\ \nonumber &-&2\kappa^{2}\frac{V''}{(V')^{2}}(\dot{H}+3H^{2})^{2})\langle \phi^{2}\rangle,\label{2}
\end{eqnarray}
with $\kappa\,=\,\frac{M^{2}_{Pl}}{8\pi}$.

The above results are valid for all cosmological eras. With this in mind, 
we now turn an eye to the post-inflation universe and see what the above 
implies about its subsequent evolution.

In what follows, we take the scalar field potential to be 
$\lambda \varphi^{4}$. 
As was shown in \cite{Shtanov:1994ce}, the equation of state of the 
inflaton after reheating is that of radiation, which implies 
$\varphi(t)\varpropto 1/a(t)$.

\section{The Radiation Epoch}

The radiation epoch followed on the heels of inflation. The EMT in this 
case reads
\be
\tau_{00} \, = \,(\frac{1}{16}\kappa^{2}\frac{V''}{(V')^{2}}\frac{1}{t^{4}}-\frac{\kappa}{t^{2}}) \langle \phi^{2} \rangle,
\ee
\be
\tau_{ij} \, = \,a^{2}(t)\delta_{ij}(-3\frac{\kappa^{2}}{t^{2}}+4(\dot{\varphi})^{2}-\frac{1}{16}\kappa^{2}\frac{V''}{(V')^{2}}\frac{1}{t^{4}})\langle \phi^{2}\rangle .
\ee

The first thing we notice is that, if the time dependence of 
$\langle \phi^{2} \rangle$ is negligible, the EMT acts as a tracker with 
every term scaling as $1/a^{4}(t)$ (except for the $\dot{\varphi}$ which 
scales faster and which we ignore from now on).

We now determine the time dependence of $\langle \phi^{2} \rangle$, where
\be
\langle \phi^{2} \rangle\,=\,\frac{\psi^{2}}{V}\int{\,d^{3}\v{x}\,\,d^{3}\v{k_{1}}\,d^{3}\v{k_{2}}}\,f(\v{k_{1}})f(\v{k_{2}})e^{i(\v{k_{1}}+\v{k_{2}})\cdot \v{x}},
\ee
with
\be
f(\v{k}) \, = \,
\sqrt{V}(\frac{k}{k_{n}})^{-3/2-\xi}k^{-3/2}_{n}e^{i\alpha(\v{k})}.
\ee\label{integral}
Here, $\psi$ represents the amplitude of the perturbations (which is
constant in time), $\xi$ 
represents the deviation from a Harrison-Zel'dovich spectrum, 
$\alpha(\v{k})$ is a random variable, and $k_{n}$ is a normalization scale.

Taking $\frac{\Lambda}{a(t)}$ as a time-dependent, infra-red cutoff and 
the Hubble scale as our ultra-violet cutoff, and focusing on the case of a 
nearly scale-invariant spectrum, the above simplifies to
\begin{eqnarray} \label{evalue}
\langle \phi^{2} \rangle\,&=&\, 4\pi\psi^{2}k^{-2\xi}_{n}\int^{H}_{\frac{\Lambda}{a(t)}}{}dk_{1}\frac{1}{k_{1}^{1-2\xi}}\\ 
\end{eqnarray}
In the limit of small $\xi$, the above reduces to 
\be
\langle \phi^{2} \rangle\,\cong\,4\pi\psi^{2}\ln(\frac{a(t)H}{\Lambda}).
\ee
The time variation of the above quantity is only logarithmic in time
and hence not important for our purposes. As well, given the small 
amplitude of the perturbations, $\langle \phi^{2} \rangle \ll 1$. Note that 
this condition is opposite to what needs to happen in the scenario 
when gravitational back-reaction ends inflation. 

Now that we have established that the EMT acts as a tracker in this epoch, 
we still have to determine the magnitude of $\tau_{00}$ and the 
corresponding equation of state. In order to do this, as in 
\cite{Shtanov:1994ce}, we assume that the preheating temperature is 
$T=10^{12}$GeV, the quartic coupling $\lambda=10^{-12}$, and the 
inflaton amplitude following preheating is $\varphi_{0}=10^{-4}M_{Pl}$. 
Making use of 
\be
a(t)\,=\,(\frac{32\pi\rho_{0}}{3M^{2}_{Pl}})^{1/4}t^{1/2},
\ee 
where $\rho_{0}$ is the initial energy density of radiation, we find
\be
\tau_{00}\,=-\kappa(\frac{32\pi\rho_{0}}{3M^{2}_{Pl}})\frac{1}{a^{4}(t)}[1-\frac{1}{8}]\langle \phi^{2} \rangle\,\cong\,-\frac{4}{3}\frac{\rho_{0}}{a^{4}(t)}\langle \phi^{2} \rangle,
\ee
\be
\tau_{ij}\,=\,-a^{2}(t)\delta_{ij}\kappa(\frac{32\pi\rho_{0}}{3M^{2}_{Pl}})\frac{1}{a^{4}(t)}[3+\frac{1}{8}]\langle \phi^{2} \rangle\,\cong\,-4\frac{\rho_{0}}{a^{4}(t)}\langle \phi^{2} \rangle.
\ee

We find that, as in the case of an inflationary background, the energy 
density is negative. However, unlike during inflation, the equation of state 
is no longer that of a cosmological constant. Rather, $w\,\cong\,3$. 
Clearly, due to the presence of $\langle \phi^{2}\rangle$, this energy 
density is sub-dominant. Using the value of $\psi$ in (\ref{evalue})
determined by the normalization of the power spectrum
of linear fluctuations from CMB experiments \cite{COBE}, we can estimate 
the magnitude to be approximately four orders of magnitude below that of 
the cosmic fluid. Any observational constraints that could arise during 
the radiation era (e.g. from primordial nucleosynthesis, or the CMB) will 
hence be satisfied.

\section{Matter Domination}

During the period of matter domination, we find that the EMT reduces to
\be
\tau_{00}\,=\,(\frac{2}{3}\frac{\kappa^{2}}{\lambda}\frac{a^{4}(t)}{\varphi^{4}}\frac{1}{t^{4}}-\frac{8}{3}\frac{1}{t^{2}})\langle \phi^{2} \rangle.
\ee
\be
\tau_{ij}\,=\,(-\frac{2}{3}\frac{\kappa^{2}}{\lambda}\frac{a^{4}(t)}{\varphi^{4}}\frac{1}{t^{4}}-\frac{8}{3}\frac{1}{t^{2}})\langle \phi^{2} \rangle.
\ee
In arriving at these equations, we are assuming that the matter fluctuations
are carried by the same field $\varphi$ (possibly the inflaton)
as in the radiation epoch, a field
which scales in time as $a^{-1}(t)$ 
\footnote{Even if we were to add a second scalar
field to represent the dominant matter and add a corresponding second
matter term in the constraint equation (\ref{constr}), it can be seen that
the extra terms in the equations for the effective EMT decrease in time
faster than the dominant term discussed here.}. 
This result is quite different from what was obtained in the radiation era 
for the following reason: previously, we found that both terms in 
$\tau_{00}$ scaled in time the same way. Now, we find (schematically)
\be
\tau_{00} \propto \frac{\kappa^{2}}{a^{2}(t)}-\frac{\kappa}{a^{3}(t)}.
\ee

The consequences of this are clear: the first term will rapidly come to 
dominate over the second, which is of approximately the same magnitude at 
matter-radiation equality. This will engender a change of sign for the 
energy density and cause it to eventually overtake that of the cosmic fluid.  
The same scaling behaviour is present in $\tau_{ij}$ and so the equation of 
state of the EMT will rapidly converge to that of a cosmological constant,
but this time one corresponding to a positive energy density.

Matter-radiation equality occurred at a redshift of about 
$z \approx 10^4$ and we find that 
\be
\tau_{00}(z=0)\,\simeq\,\rho_{m}(z=0),\quad w\,\simeq\,-1, 
\ee
and thus we are naturally led to a resolution of the both aspects of
the dark energy problem. We have an explanation for the presence of
a source of late-time acceleration, and a natural solution of the
``coincidence'' problem: the fact that dark energy is rearing its
head at the present time is directly tied to the observationally
determined normalization of the spectrum of cosmological perturbations.

\section{Dark Energy Domination and Inflaton Back-reaction}

Does this model predict that, after an initial stage of matter domination, 
the universe becomes perpetually dominated by dark energy? To answer this 
question, one needs to examine the effects of back-reaction on the late time 
scalar field dynamics.

The EMT predicts an effective potential for $\varphi$ that differs from the 
simple form we have been considering so far. During slow-roll, we have that
\be
V_{eff} \, = \,V+\tau_{0}^{0}.\label{effective potential}
\ee
One might expect that this would lead to a change in the spectral index of 
the power spectrum or the amplitude of the fluctuations. To show that this 
is not the case, we can explicitly calculate the form of $V_{eff}$ for the 
case of an arbitrary polynomial potential and see that, neglecting any 
$\varphi$ dependence of $\langle\phi^{2}\rangle$, (\ref{effective potential}) 
implies an (a priori small) renormalization of the scalar field coupling. We 
find that the inclusion of back-reaction does not lead to any change in the 
spectral index (in agreement with \cite{Martineau:2005aa}) or to any 
significant change in the amplitude of the perturbations.

During radiation domination, we find that the ratio of 
$\frac{\tau_{0}^{0}}{V}$ is fixed and small, so that scalar field 
back-reaction does not play a significant role in this epoch.
In fact, back-reaction on the scalar field does not become important 
until back-reaction begins to dominate the cosmic energy budget. In that case, 
\be
V_{eff} \sim \frac{1}{\varphi^{4}},
\ee
causing the $\varphi$ to ``roll up'' it's potential. Once $\varphi$ comes 
to dominate, the form of the effective potential changes to
\be
V_{eff}\sim \varphi^{4},
\ee
and $\varphi$ immediately rolls down it's potential, without the benefit 
of a large damping term (given by the Hubble scale).

Thus, this model predicts alternating periods of dark energy/matter 
domination, which recalls the ideas put forth in \cite{Brandenberger:1999su}.

From the point of view of perturbation theory, we see that in the regime where the higher-order terms begin to dominate and the series would be expected to diverge, these corrections are then suppressed and become sub-dominant again.

\section{Discussion and Conclusions}

To recap, we find that, in the context of inflationary cosmology, the EMT of
long wavelength cosmological perturbations can provide a candidate for
dark energy which resolves the ``new cosmological constant'' (or
``coincidence'' problem in a natural way. Key to the success of
the mechanism is the fact that the EMT acts as a tracker  during the period of radiation domination, but redshifts
less rapidly than matter in the matter era. The fact that our dark energy
candidate is beginning to dominate today, at a redshift $10^4$ later than
at the time of equal matter and radiation is related to the observed 
amplitude of the spectrum of cosmological perturbations.

We wish to conclude by putting our work in the context of other recent
work on the gravitational back-reaction of cosmological perturbations.
We are making use of non-gradient terms in the EMT (as was done in
\cite{Mukhanov:1996ak,Abramo:1997hu}). As was first realized by Unruh
\cite{Unruh} and then confirmed in more detail in \cite{Ghazal1,AW1},
in the absence of entropy fluctuations, the effects of these terms
are not locally measurable (they can be undone by a local time
reparametrization). It is important to calculate the effects of
back-reaction on local observables measuring the expansion history.
It was then shown \cite{Ghazal2} (see also \cite{AW2}) that 
in the presence of entropy fluctuations, back-reaction of the non-gradient
terms is physically measurable, in contrast to the statements
recently made in \cite{Wald} \footnote{There are a number of problems present in the arguments of \cite{Wald}, in addition to this point. We are currently preparing a response that addresses the criticisms of these authors. See \cite{us}.}. In our case, we are making use of
fluctuations of the scalar field $\varphi$ at late times. As long as
this fluctuation is associated with an isocurvature mode, the
effects computed in this paper using the EMT approach should also
be seen by local observers. 

Our approach of explaining dark energy in terms of back-reaction is different 
from the proposal of \cite{Kolb:2005me}. In that approach, use is made of the 
leading gradient terms in the EMT. However, it has subsequently been shown 
\cite{Ghazal3} that these terms act as spatial curvature and that hence their 
magnitude is tightly constrained by observations. Other criticism was raised 
in \cite{Seljak} where it was claimed that, in the absence of a bare 
cosmological constant, it is not possible to obtain a cosmology which changes 
from deceleration to acceleration by means of back-reaction. This criticism is 
also relevant for our work. However, as pointed out
in \cite{Buchert}, there are subtleties when dealing with spatially averaged 
quantities, even if the spatial averaging is over a limited domain, and that 
the conclusions of \cite{Seljak} may not apply to the quantities we are 
interested in.

There have also been attempts to obtain dark energy from the back-reaction of
short wavelength modes \cite{Rasanen,Alessio,Kolb2}. In these approaches, 
however, nonlinear effects are invoked to provide the required magnitude of the
back-reaction effects.

We now consider some general objections which have been raised regarding
the issue of whether super-Hubble-scale fluctuations can induce
locally measurable back-reaction effects. The first, and easiest to refute, 
is the issue of causality. Our formalism is based entirely on the equations of 
general relativity, which are generally covariant and thus have causality 
built into them. We are studying the effects of super-Hubble but sub-horizon 
fluctuations \footnote{We remind the reader that it is exactly because 
inflation exponentially expands the horizon compared to the Hubble radius 
that the inflationary paradigm can
create a causal mechanism for the origin of structure in the universe.
In our back-reaction work, we are using modes which, like those which we now
observe in the CMB, were created inside the Hubble radius during the early 
stages of inflation, but have not yet re-entered the Hubble radius in the 
post-inflationary period.}. Another issue is locality. As shown in \cite{Lam}, 
back-reaction effects such as those discussed here can be viewed in terms of 
completely local cosmological equations. For a more extensive discussion, the reader is referred to \cite{us}.

In conclusion, we have presented a model which can solve the dark energy 
problem without resorting to new scalar fields, making use only of 
conventional gravitational physics. The effect of the back-reaction of the 
super-Hubble modes is summarized in the form of an effective energy-momentum 
tensor which displays distinct behaviour during different cosmological epochs. 

\begin{acknowledgments}

This work is supported by funds from McGill University,
by an NSERC Discovery Grant and by the Canada Research Chair program. P.M. would like to thank Mark Alford and the Washington University physics department for their hospitality while part of this work was being completed.

\end{acknowledgments}

\end{document}